\documentclass[journal]{IEEEtran}
\usepackage{amsmath,amssymb,amsfonts,tabularx}
\usepackage[hidelinks]{hyperref}
\usepackage{cite}
\usepackage{url}
\usepackage{paralist}
\usepackage{graphicx}
\usepackage{makecell}
\usepackage{xcolor}
\usepackage{array,multirow}
\usepackage{xspace}
\usepackage{flushend}
\graphicspath{{figures/}}

\def\name{\textsc{Dolos}\xspace}

\newcommand{\bpar}[1]{\par\noindent\textbf{#1}}
\newcommand{\tabitem}{~~\llap{\textbullet}~~}

\begin{document}
\title{\name: A Novel Architecture for Moving Target Defense}

\author{
    \IEEEauthorblockN{
        Giulio Pagnotta\IEEEauthorrefmark{1},
        Fabio De Gaspari\IEEEauthorrefmark{1},
        Dorjan Hitaj\IEEEauthorrefmark{1},
        Mauro Andreolini\IEEEauthorrefmark{2},
        Michele Colajanni\IEEEauthorrefmark{3},
        Luigi V. Mancini\IEEEauthorrefmark{1}\\
    }
    \IEEEauthorblockA{
        \IEEEauthorrefmark{1}Sapienza University of Rome, 
        \{pagnotta, degaspari, hitaj.d, mancini\}@di.uniroma1.it\\
    }
    \IEEEauthorblockA{
        \IEEEauthorrefmark{2}University of Modena and Reggio Emilia, 
        mauro.andreolini@unimore.it\\
    }
    \IEEEauthorblockA{
        \IEEEauthorrefmark{3}University of Bologna,
        michele.colajanni@unibo.it
    }
}

\maketitle

\begin{abstract}
Moving Target Defense and Cyber Deception emerged in recent years as two key proactive cyber defense approaches, contrasting with the static nature of the traditional reactive cyber defense. The key insight behind these approaches is to impose an asymmetric disadvantage for the attacker by using deception and randomization techniques to create a dynamic attack surface. 
Moving Target Defense (MTD) typically relies on system randomization and diversification, while Cyber Deception is based on decoy nodes and fake systems to deceive attackers. 
However, current Moving Target Defense techniques are complex to manage and can introduce high overheads, while Cyber Deception nodes are easily recognized and avoided by adversaries.

This paper presents \name, a novel architecture that unifies Cyber Deception and Moving Target Defense approaches. \name is motivated by the insight that deceptive techniques are much more powerful when integrated into production systems rather than deployed alongside them. \name combines typical Moving Target Defense techniques, such as randomization, diversity, and redundancy, with cyber deception and seamlessly integrates them into production systems through multiple layers of isolation. 
We extensively evaluate \name against a wide range of attackers, ranging from automated malware to professional penetration testers, and show that \name is effective in slowing down attacks and protecting the integrity of production systems. We also provide valuable insights and considerations for the future development of MTD techniques based on our findings.

\end{abstract}

\begin{IEEEkeywords}
Moving Target Defense, Active Defense, Cyber Deception, Intrusion Detection.
\end{IEEEkeywords}
\IEEEpeerreviewmaketitle

\section{Introduction}
\IEEEPARstart{T}{raditional} cyber defense techniques are typically based on a detection-reaction paradigm where various mechanisms are deployed to identify early signs of intrusion. Upon detection, intruders are generally blocked, and relevant alerts are forwarded to the Computer Security Incident Response Team (CSIRT) for remediation. This \emph{reactive} approach to cybersecurity has considerably evolved over the past decades, and the core concept of detection-reaction is embedded in the large majority of cyber defense tools available today. However, reactive approaches create a disadvantage for defenders because of the asymmetrical relationship between attackers and defenders. Indeed, while reactive cyber defenses are static and come from fixed procedures, they must be effective 

even against novel and unfamiliar adversaries using constantly changing attack methods and exploiting new vulnerabilities.

Moving Target Defense (MTD)\cite{jajodia2011moving} (also known as Active Defense) and Cyber Deception~\cite{jajodia2016cyber} emerged in recent years as proactive cyber defense techniques aimed at protecting computer systems from intrusion. The key concept behind these approaches is to impose an asymmetric disadvantage for the attacker by exploiting randomization and deception techniques, creating a dynamic attack surface that is hard for the attacker to identify and exploit~\cite{cho2020toward,lei2018moving,cai2016moving,degaspari2016ahead}.

MTD-based techniques are typically based on randomization, diversification, and redundancy and aim at changing system configuration unpredictably to hamper reconnaissance and exploitation attempts~\cite{8455956,carroll2014analysis,taguinod2015toward,huang2011introducing}. 
Cyber Deception techniques generally rely on mock-ups of real systems and services deployed on separate, fake nodes in the network with the goal of misleading attackers and attracting them away from sensitive targets~\cite{ferguson2021examining,mairh2011honeypot,qin2023hybrid}.

However, these approaches suffer several limitations. MTD approaches, particularly randomization-based ones, can impose large overheads on the system they are deployed on, are complex to implement, and risk affecting legitimate services~\cite{cho2020toward}. Cyber Deception techniques do not hinder legitimate services and pose no overhead on production systems, as they are typically deployed on separate nodes or machines. However, it has been shown that expert adversaries can easily recognize and bypass them to focus on the actual production systems~\cite{mukkamala2007detection,holz2005detecting}.

This paper presents \name, a new architecture to unify Moving Target Defense and Cyber Deception. \name is designed to seamlessly integrate into production systems to provide deception and MTD capabilities through the orchestration of fake services and randomization of the attack surface. Unlike typical MTD, \name \emph{does not randomize real services}: only deceptive services and system properties are manipulated. This approach allows \name to reduce overhead and management complexity compared to previous MTD techniques. On the other hand, \name integrates directly in production systems through multiple layers of isolation, making any bypass much harder compared to typical Cyber Deception techniques. 
The concept behind \name is that deceptive techniques are much more powerful when integrated into the production systems rather than deployed alongside them.
We show that embedding \name in production systems can lead attackers to completely \emph{disregard} them, mistaking them for fake machines. 
We evaluate \name against different types of attackers, ranging from automated malware to professional penetration testers, and unequivocally show its effectiveness in thwarting attacks. 

In this paper, which builds and expands upon our previous work of~\cite{degaspari2016ahead}, we make the following new contributions:

\begin{compactitem}
 \item We present \name, a novel architecture to unify Cyber Deception and MTD. \name integrates deception and randomization capabilities directly into production systems, providing the advantages of both MTD and Cyber Deception techniques, without the drawbacks.
 \item We provide what is, to the best of our knowledge, the most thorough experimental evaluation of an MTD approach to date. We study the effectiveness of \name against a wide range of different attackers, from automated malware to professional pentesters, both in the real world and virtual networks.
 \item We show that \name's approach is highly effective in slowing down attackers and protecting the integrity of production systems, even against expert human attackers.

 \item We make available the code to reproduce our results at~\href{https://github.com/pagiux/dolos}{\url{https://github.com/pagiux/dolos}}
\end{compactitem}

In particular, we show that (1) \name can effectively trap automated malware for extended periods of time, hindering reconnaissance and making remediation easier; (2) average human attackers are unable to identify and avoid \name services, and are confused by \name MTD-deception tools; (3) expert human attackers consistently fail to compromise machines protected by \name; (4) \name significantly increases the time-to-compromise when deployed and greatly decreases attack success rate.
\section{Threat Model}
\label{sec:threat_model}

\begin{figure}
    \centering
    \includegraphics[width=0.8\columnwidth]{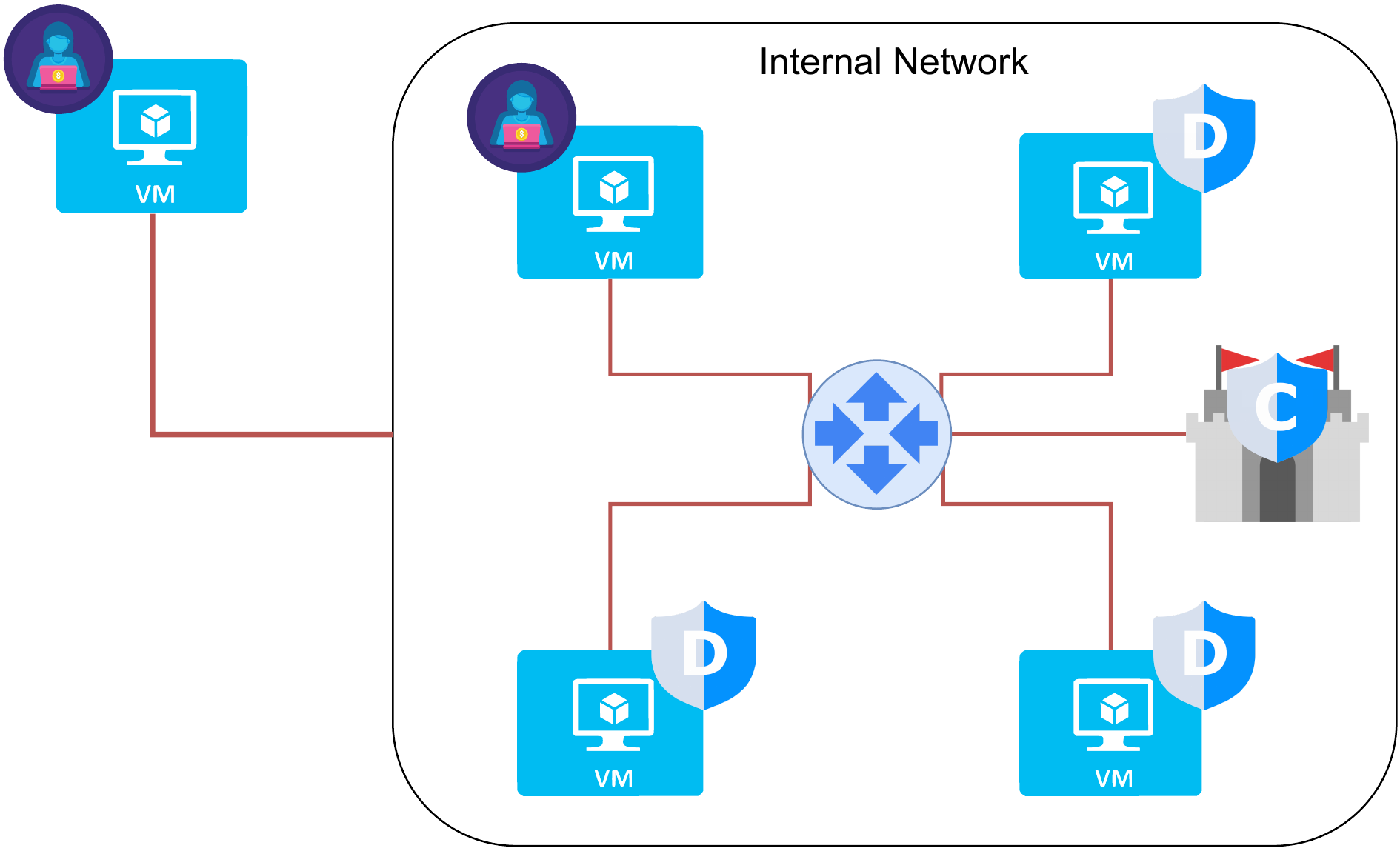}
    \caption{\name threat model. We consider both internal adversaries which reside inside the network that \name is protecting, as well as external adversaries that can only reach machines on the network perimeter. The \name controller is considered trusted.}
    \label{fig:threat_model}
\end{figure}

Figure~\ref{fig:threat_model} illustrates the threat model. We consider the setting of a computer network, such as a company network, that needs to be secured from external and internal attackers. A \name Agent (D in the figure) is deployed on each computer in the network (called \emph{production systems} from here on) and provides them with advanced deception and MTD capabilities. A single \name Controller (C in the figure)  manages all the Agents in the network;  the communication between Agent and Controller is unicast. Each Agent is managed individually by the Controller and no synchronization among the Agents or between messages from the different Agents is required. The \name Controller is considered trusted, and the communication between the Controller and the Agent is considered tamper-proof and secure, for instance by means of TLS. The \name Agent \emph{is not} considered trusted: it can be compromised by adversaries which can, for instance, use it to send forged logs and alerts to the controller. 
Similar to other MTD techniques, \name is assumed to be part of a complete set of cyber defense tools to provide defense in depth. \name is designed to slow down and thwart reconnaissance and lateral movement activities from attackers, making it more difficult to find and exploit vulnerable services and applications. \name is not designed to patch critical vulnerabilities and misconfigurations of system applications that an attacker can exploit after obtaining access to the production system. 

We consider two types of adversaries: (1) external attackers and (2) internal attackers. External attackers are located outside the perimeter of the network and can only initiate connections toward production systems that are on the network perimeter, exposed to the Internet.
External attackers are therefore limited in their interactions with production systems and can only query exposed services remotely in an attempt to find vulnerabilities and compromise them.
Internal attackers reside on a production system within the network and have full control over it. Internal attackers can reach any production system within the network.
Both internal and external attackers can scan the network to identify reachable production systems and are allowed to interact with them without limitation. We do not assume IDS or other traditional tools to prevent or limit the activity of the attacker in any way.
We assume that legitimate users are well-behaved: they only interact with the production systems through standard applications provided to them with the specific configuration given by the manager of the network, and do not attempt to perform suspicious activities such as network scanning, enumeration or access files on the system that do not belong to them. Therefore, in our model legitimate users never interact with \name Agent's services. Any interaction directed towards any service provided by the \name Agent is considered suspicious and originated by a potential adversary.

\section{\name}
\label{sec:adarch}
This section presents \name, our novel Moving Target Defense architecture. The key idea behind \name is that deception is a much more powerful tool when embedded directly in the systems it should protect rather than deployed alongside them. To this end, \name is designed to seamlessly integrate into production systems to provide advanced deception and annoyance capabilities by orchestrating fake services and randomizing system resources. The goal of \name is two-fold: on the one hand, protecting the production systems by disguising real services among multiple, real-looking fake ones; on the other, slowing down attackers by randomizing the attack surface through fake services and resources, thus allowing ample time for the CSIRT to detect and respond to the threat.

Unlike traditional MTD tools, which directly modify the real services and the configuration of production systems through randomization and diversification techniques~\cite{cho2020toward}, \name only alters fake services and system resources that are added by a \name Agent running on the production systems. The real services running on the production system are never randomized or reconfigured, therefore avoiding the management complexities and overhead introduced by traditional MTD techniques. \name can alter fake services and system resources without limitation since they have no impact on the activity of legitimate users, and no additional mechanisms are required to guarantee the consistency and availability of these services. Furthermore, given the direct integration of the fake services and resources into the production system, \name avoids the limitations of traditional cyber deception techniques whereby adversaries can easily identify deceptive systems such as honeypots and avoid them~\cite{mukkamala2007detection,holz2005detecting}. Since \name fake services are deployed on the same production systems as real services, an attacker will need to thoroughly interact with all of them to assess whether they are real or fake in order to identify potential vulnerabilities, as demonstrated in our experimental evaluation.
Finally, \name is designed to transparently integrate with and be deployed alongside other cyber defense tools to provide defense in depth to the production systems. Since \name does not affect real services, it can also be deployed alongside other traditional MTD techniques that employ service randomization and diversification without any conflict.

Overall, the benefits of integrating \name into production are the following:

\textbf{Real-Time Protection:} \name's MTD and deception modules transparently provide all production systems in the network with real-time, always-on defense capabilities. As highlighted Section~\ref{sec:eval_average}, \name modules are effective in slowing down and confusing attackers, as well as increasing the amount of interaction with the target system required for adversaries. These properties make it considerably easier to detect intrusions and provide ample time and opportunity for incident and response teams to intervene.

\textbf{Operational Continuity:} \name integration ensures minimal disruption to production system operations. As shown in Section~\ref{sec:eval_overhead}, the production system services remain fully functional and are not affected in any way by \name, which operates only on fake, deceptive services and resources. 

\textbf{Reduced Management Complexity:} \name modules deploy and interact only with fake services and resources. This reduces the complexity and overhead associated with managing multiple MTD solutions that continuously shuffle and randomize real services. Indeed, as we show in Section~\ref{sec:eval_overhead} and discuss in Section~\ref{sec:related_works}, traditional MTD tools require complex ad-hoc solutions to ensure that real services are not disrupted, increasing the burden on network management and rendering network troubleshooting more difficult.

\textbf{Adaptive Defense:} the architecture of \name is designed to allow dynamic deployment and reconfiguration of the modules running in the production systems through the \name Agent (see Section~\ref{sec:agent}). This provides \name with the capability to dynamically adapt deployed modules based on specific attack patterns or actions performed by the adversary against \name modules, making the system highly adaptable to evolving threats. An initial implementation of dynamic adaptation is currently implemented in \name MTD Modules, as discussed in Section~\ref{sec:eval_prof}. However, a completely autonomous control logic to detect attack patterns and abstract adversarial goals into concrete actions is left for future work.

\textbf{Cost-Effectiveness}: the integration of \name into the production system is more cost-effective compared to the deployment and management of multiple, separate honey-devices in the network. Organizations can leverage existing infrastructure and tools for the management of \name, maximizing their investments while enhancing security capabilities.

The remainder of this section presents the architecture of \name  in Section~\ref{sec:architecture}, the MTD tools (called \emph{MTD modules} from here on) implemented in \name in Section~\ref{sec:modules}, and an overview of the sandboxing and hardening techniques in Section~\ref{sec:hardening}.

\subsection{Architecture}
\label{sec:architecture}
The \name system consists of two main components: the \name Agent and the \name Controller. The \name Agent is the architectural component deployed on production systems and effectively implements deceptive capabilities. The \name Controller provides command and control functionality for the management of the Agent. In a real-world deployment, an Agent is installed in each production system that needs protection, with a single Controller managing all the agents in the network. In what follows, we describe the \name Agent and Controller and their functionality in detail.

\begin{figure}[t]
  \centering
  \includegraphics[width=0.8\columnwidth]{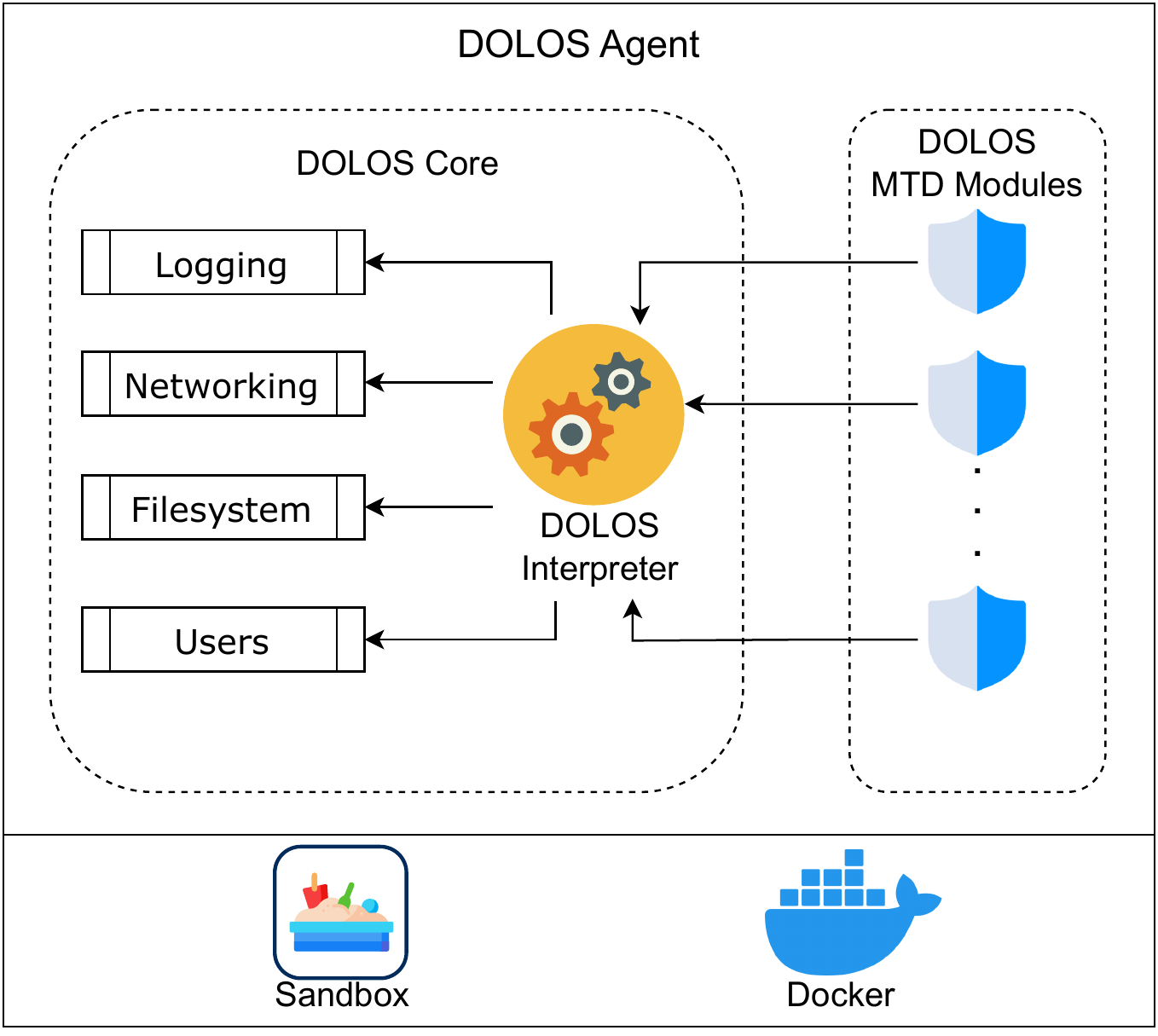}
  \caption{Architecture of the \name Agent. MTD tools are implemented as modules, executed by an interpreter in the \name Core. The Agent is isolated from the real production system through multiple layers of sandboxing.}
  \label{fig:agent_arch}
\end{figure}

\subsubsection{\name Agent}
\label{sec:agent}

The \name Agent is implemented as a multi-layered architecture based on a clear separation between different component modules. Figure~\ref{fig:agent_arch} presents an overview of the architecture. 
The \name Core component of the Agent is a software module written in C that integrates a Python interpreter to facilitate the prototyping and development of MTD tools. The C component of the core efficiently implements a set of advanced functionalities typically required by MTD modules and exposes an interface to access them. This interface is exposed through the integrated Python interpreter that is extended to interact with the C components of the Core seamlessly. 
The extension of the interpreter was carefully designed to accommodate the concurrent nature of the \name Core. In particular, the explicit management of the internal reference count of Python objects to periodically clean up memory and the fine-grained handling of the Global Interpreter Lock (GIL) posed non-trivial technical challenges.

The C component of the \name Core implements standardized functions for logging, networking, filesystem, and user-handling-related operations.
In particular, the logging component exposes an interface to standardize the generation, storage, and management of logs across all the MTD modules, as well as the functionalities to send the event logs to the controller. The networking component provides methods to seamlessly send and receive network connections concurrently, relying on an implementation of a thread pool within the Core so that MTD modules can be developed without consideration for these complexities. The filesystem component provides a system-agnostic abstraction layer that makes the integration of MTD modules that operate on files easier. It provides interfaces for file creation, integrity checking, and deployment of monitors to detect file and directory modifications. Finally, the user-handling component exposes common methods for the creation and handling of user profiles, such as the setup of user directory trees and of various profile properties.

We highlight that the \name Core component can be easily extended to integrate additional functionalities without any particular limitation, except for the boundaries imposed by the isolation of the \name Agent and the underlying production system.
Furthermore, as a result of the integration of the Python interpreter, a Python interface to these functionalities can be easily provided to MTD module developers.

MTD modules are implemented as external components and are executed by the \name interpreter. This modular design of the Agent allows for easy extensibility and avoids duplication of functionality between different MTD modules. The Agent is separated from the underlying production system by multiple layers of isolation through container-based techniques and sandboxing. The sandboxing layer is carefully designed to limit the Agent's access only to the necessary functionality of the underlying system and to reduce the overhead for the production system. As our experimental analysis in Section~\ref{sec:eval_overhead} shows, our design allows \name to have negligible impact on system performance. We provide more details on the isolation techniques used in Section~\ref{sec:hardening} and a detailed quantitative analysis of the performance overhead in Section~\ref{sec:eval_overhead}.

The main advantages of the \name Agent over previous container-based solutions are its modularity, extensibility, and efficiency. Typically, a basic container is used to run multiple independent MTD tools simultaneously. The container-based approach was used mainly for ease of deployment and to provide some level of isolation from the underlying system~\cite{degaspari2016ahead}. On the other hand, the \name Agent implements standard functions in the Core component and exposes them to the MTD modules through a unified interface. This modular approach reduces overall resource utilization, as different modules do not need to instantiate the same resource multiple times. The centralization of common functions in the Core component also allows for fine-grained control over the interactions between the Agent and the underlying system, which occur only through the Core libraries. Finally, new MTD modules can be integrated into the Agent in a much easier manner than in typical container-based approaches. 

\subsubsection{\name Controller}
\begin{figure}[t]
  \centering
  \includegraphics[width=0.7\columnwidth]{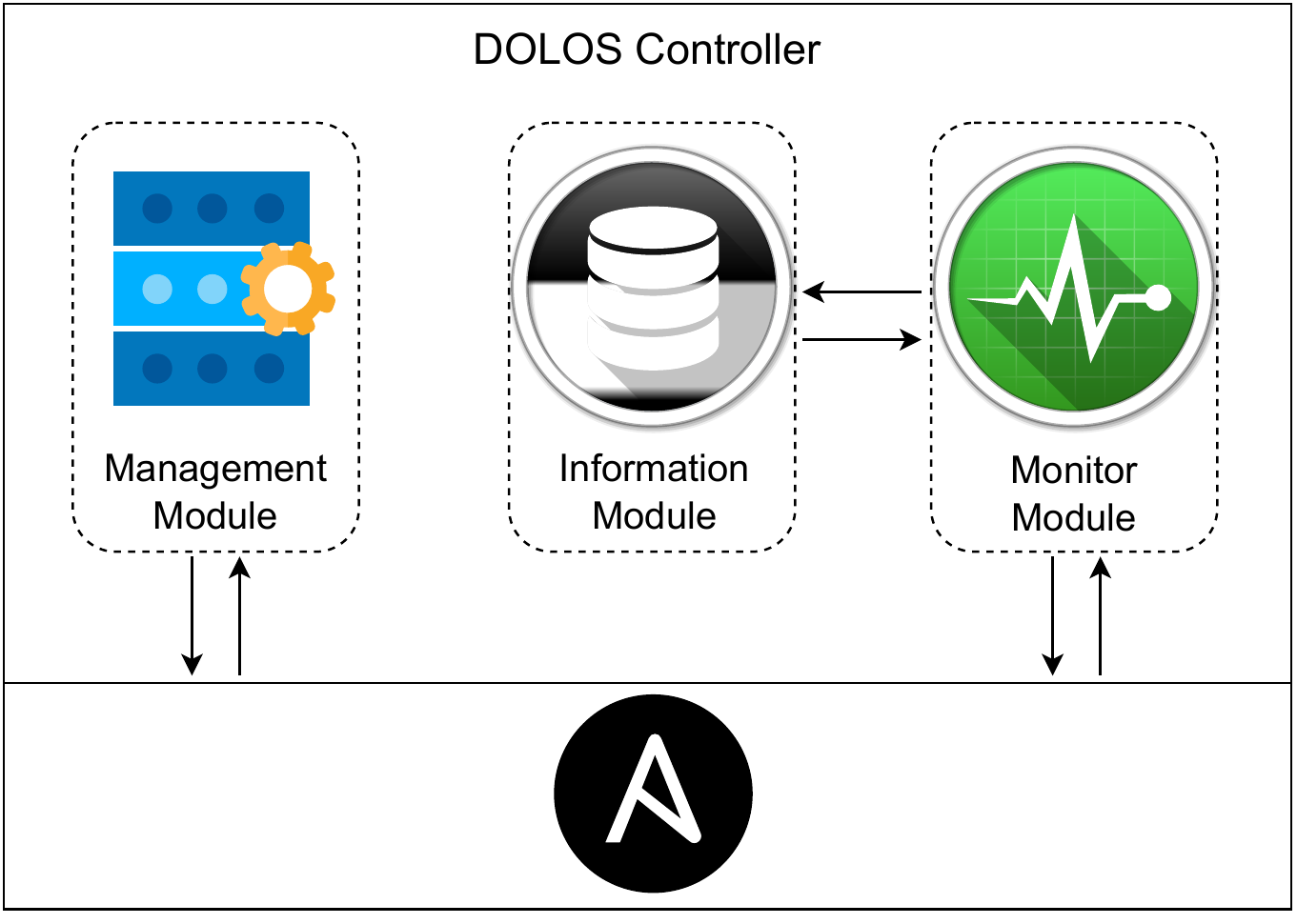}
  \caption{Architecture of the \name Controller.}
  \label{fig:controller_arch}
\end{figure}

The architecture of the \name Controller is presented in Figure~\ref{fig:controller_arch}. The Controller is the key component of \name that provides management, monitoring, and control functions. It comprises three main modules: Management, Information, and Monitor. The Management module offers a remote management interface for the Controller and, indirectly, for the Agents. All control actions such as installing Agents, deploying new MTD modules, or restarting an Agent are executed by this component and are carried out through an interface layer implemented through the Ansible orchestrator~\cite{ansible}. The Information module maintains status updates, processed logs from MTD modules, and alerts generated by the Agent. The storage of logs generated by MTD modules is currently implemented in the Structured Threat Information Expression (STIX) format~\cite{stix}, thus allowing easier processing of security events and implementation of automated decision-making. Finally, the Monitor module is a correlation engine that regularly processes data in the Information module and generates alerts on potential threats detected. In the present version, response and remediation actions are not automated by the Controller and must be undertaken by the Incident Response Team. We plan to introduce autonomous learning capabilities in the Controller in future work, as detailed in Section~\ref{sec:discussion}.

\subsection{\name MTD Modules}
\label{sec:modules}
This section describes the current capabilities of \name and the MTD modules currently implemented by the Agent. We subdivide \name MTD modules into three categories: networking-related, filesystem-related, and user account-related.

\subsubsection{Networking Modules} The Networking modules provide a set of essential functionalities to spoof well-known services and fake open ports. The main goal of these modules is to make it hard for an adversary to find real services and to slow down the reconnaissance phase of the attack significantly. Currently, \name implements four networking MTD modules:
\begin{itemize}
    \item Portspoof, a module that fakes the presence of arbitrary services behind a set of selected open ports. It emulates valid services by dynamically generating signatures to respond to service probes.
    \item Honeyports, a module that attaches and listens on a set of predefined ports, completing any incoming connection requests. Any IP address that initiates and completes a connection with Honeyports is added to a blocklist.
    \item Invisiport, a module that implements deceptive blocklisting of incoming connections. It listens for probes on a set of fake, open ports and adds the initiator to an internal blocklist when a connection is detected. All connections from blocklisted initiators are refused on all ports, except for a second set of fake Invisiport services that will still be reachable.
    \item Endlessh, a module that implements an SSH tarpit to slow down bruteforce attacks. At each connection request, before sending its SSH identification string, it slowly sends a series of banners keeping the bruteforcing script locked up for a long time.
    \item Artillery, a module that implements SSH bruteforce monitoring. It analyzes login attempts on any SSH shells it is configured to listen on, including potentially fake \name shells, and reports suspicious chains of failed login attempts.
\end{itemize}

\subsubsection{Filesystem Modules} The Filesystem modules implement file-based deceptive functionalities. The main goal of these modules is to provide early detection of local access breaches and the ability to terminate access to compromised user accounts. Currently, \name implements the following filesystem MTD modules:
\begin{itemize}
    \item Honeyfiles, a module that monitors file and directory access in the production system. It attaches a listener to a set of specified files and directories and monitors whether unexpected actions are performed. It can also apply different countermeasures, such as terminating the offending process or locking out users.
    \item Cryptolocked, a module that deploys fake files in the production system called trip files. It attaches a listener to all deployed trip files and triggers an alert whenever access is detected. 
\end{itemize}

\subsubsection{User Account Modules} The User Account modules implement MTD functionality related to fake user accounts. The main goal of these modules is to provide the attacker with a believable environment to interact with in order to slow him down. Currently, \name implements one user account MTD module:
\begin{itemize}
    \item Honey Account, a module that deploys fake, realistic user accounts in the system. Each account is initialized with a semi-randomized directory structure and a set of files. Moreover, a high-interaction faux shell is attached to each honey account to provide realistic interactions with adversaries. The honey accounts are accessible through a regular SSH server. 
\end{itemize}

\subsection{\name Hardening}
\label{sec:hardening}
Introducing additional services and open ports in production systems, even when fake, inevitably increases the attack surface of the systems. Potentially, an attacker could compromise a \name MTD module to obtain local access to the system and escalate privileges to fully compromise the machine. \name employs a multi-layer sandbox architecture to limit this possibility and reduces the privileges of the Agent to those strictly necessary. The first layer of isolation used in the \name Agent is based on container technology~\cite{docker}. \name processes are isolated inside a container from normal production system processes and cannot see nor interact with them in any manner. \name Agents also have their own network stack, and interaction with the network interfaces of the production system is allowed only through typical network connections. This prevents the \name Agent from sniffing traffic directed to the production system. Beyond the isolation provided by the container, \name deploys additional tools to limit the Agent's access to the underlying system. Access to the production system kernel is filtered through an allowlist, authorizing execution only of specific system calls. The interaction between the Agent and the production system is further limited through mandatory access control (MAC)~\cite{apparmor}. \name uses strict MAC configuration policies that prevent the Agent from accessing any object it does not strictly need. Finally, \name also uses fine-grained capabilities configuration to limit container permissions only to those effectively required by the system, which is mainly the use of raw sockets~\cite{raw_socket}. To use the \name Agent as an entry point in the system, an attacker would have to: (1) compromise a \name MTD module; (2) escalate privileges within the Agent's container; (3) escape the container through a vulnerability; (4) achieve all this while being limited by \name's system call allowlist, MAC and capability limitation system.

Finally, an attacker that can compromise a \name Agent but is unable to escape the isolation to the underlying system can only use the Agent to communicate with the Controller. However, this is inconsequential, as the Agent does not have the ability to send any commands or carry out any actions for the attacker. The attacker can either prevent local alerts and log forwarding to the Controller or send fake logs/alerts. Preventing log forwarding after compromising the Agent is ineffective because the attacker has already interacted with the MTD modules to gain access, and all relevant logs and alerts have already been sent to the Controller. Meanwhile, sending forged logs or alerts would serve no real purpose, as it may only risk triggering an alert from the Controller to the CSIRT. These alerts can be further validated by correlating data from the \name Agent with those from other traditional defense systems in the network.

\section{Evaluation}
This section evaluates \name across a wide range of conditions and 
against four different types of adversaries: automated malware, average human attackers, expert human attackers, and professional human attackers. Finally, we study the overhead introduced by the \name framework on the production systems.

\subsection{Automated Adversary: Malware}
This section evaluates the ability of \name to slow down automated attacks by malware. We design two different settings to analyze malware interactions with our MTD tools: (I) focusing on interactions at the filesystem level, representing the case of malware running on a local machine in the network, for instance, after a successful phishing campaign; and (II) focusing on interactions at the network level, representing the case of an external attacker trying to get a foothold in the network. With these experiments, we aim to answer the following research questions: (a) are file system MTD tools practical to slow down automated malware? (b) are networking MTD tools effective in thwarting and detecting malware lateral movement? (c) can networking MTD tools slow down reconnaissance and local access attempts by malware?

\begin{figure}
    \centering
    \includegraphics[width=0.9\columnwidth]{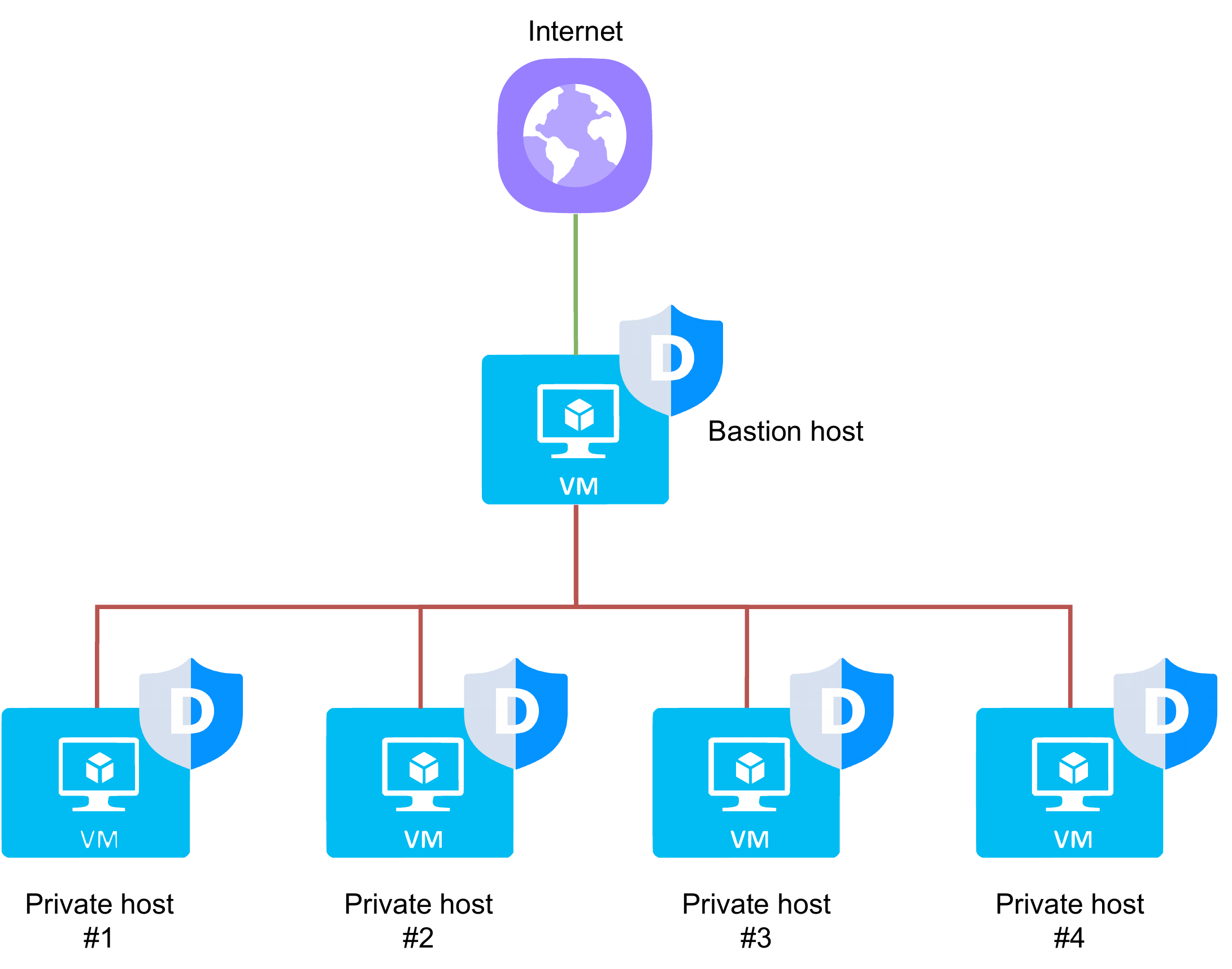}
    \caption{Experimental setup used in Setting (I) of the Automated Adversary evaluation.}
    \label{fig:malware_setup_A}
\end{figure}

\begin{table*}[t]
	\footnotesize
	\def\arraystretch{1.3}
	\centering
	\renewcommand{\tabcolsep}{3pt}
	\caption{Filesystem-level activity detected through log analysis after compromise of bastion host.}
	\begin{tabular}{|l|l|l|l|}
		\hline
		\textbf{Botnet} & \textbf{Type} & \textbf{Prevalence}&\textbf{Filesystem Activities} \\ \hline
		Unidentified \#1 & crypto miner masquerading as dhcp daemon & $7/14$ VMs & \makecell[l]{
		    \tabitem creation of \url{~/}.dhpcd executable with XMRig cryptominer. \\
		    \tabitem creation of \url{~/}.ssh/authorized\_keys with fixed key. \\
		    \tabitem creation of \url{~/}.core file containing bytecode. \\
		    \tabitem crontab launching dhpcd every 60 seconds.
		    } \\ \hline
		Muhstik	& botnet - cryptominer & $2/14$ VMs & \makecell[l]{
		    \tabitem wget downloads of multiple files to home, /tmp. \\
		    \tabitem renaming of several files and download of shell scripts.
		    } \\ \hline
		Outlaw & botnet - cryptominer & $2/14$ VMs & \makecell[l]{
		    \tabitem creation of \url{~/}.configrc and subdirectories ``a'' and ``b''. \\
		    \tabitem unpacking of XMRig cryptominer in ``a''.
		    } \\ \hline
		Unidentified \#2 & cryptominer & $1/14$ VMs & \makecell[l]{
		    \tabitem creation of multiple files in /tmp directory. \\
		    \tabitem all created files are deleted shortly after.
		    } \\ \hline
		Unidentified \#3 & cryptominer & $1/14$ VMs & \makecell[l]{
		    \tabitem wget download of script file in /home. \\
		    \tabitem script file deleted shortly after.
		    } \\ \hline
	\end{tabular}
	\label{tab:crypto_activities}
\end{table*}

\subsubsection{Experimental Setup - Setting (I)}
Figure \ref{fig:malware_setup_A} illustrates the experimental setup. We deploy five Debian VMs on the open Internet through the Google Cloud Platform Compute Engine (GCP). One of these machines acts as a bastion host, providing connectivity to the subnetwork and preventing direct connections from the Internet to the remaining four private hosts. On the bastion host, we deploy three separate docker containers: 
\begin{itemize}
    \item \name container: runs the Cryptolocked and Honeyfiles modules to deploy trip-files throughout the filesystem and to monitor file events in the \textit{/home, /data} and \textit{/tmp} directories of the SSH users and Apache container.
    \item Apache container: runs an HTTP web server exposed on port 80 comprised of a single CGI webpage. The page allows users to interact with the underlying systems and send commands to a bash shell vulnerable to the Shellshock vulnerability \cite{shellshock}.
    \item Ubuntu container: runs an SSH service on the default port 22, exposing two accounts with extremely weak username-password pairs.
\end{itemize}

On the four private hosts, we deploy \name with Honeyports that expose fake services on well-known open ports to detect any potential malware traffic generated from the bastion host. This configuration of \name allows us to evaluate both the effectiveness of file-based deception against automated malware and the usefulness of network-level MTD against lateral movements by automated adversaries.

\subsubsection{Results - Setting (I)}
We have deployed our virtual network on the open Internet for a period of 30 days, over which we analyzed all interactions with the containers in the bastion host and private hosts. Over this period, the bastion host was compromised within a few hours of deployment a total of $14$ times, with a maximum deployment-to-compromise time of 27 hours. After each compromise, the bastion host was manually terminated and redeployed. All security breaches happened through the exploitation of the weak credentials on the SSH service, while no successful attacks were detected on the HTTP Web server. 

Once local access to the bastion host was obtained, none of the malware interacted with any trip file or performed any operation in the directories monitored by \name. Furthermore, no outbound connection to any of the private hosts was detected, and in general, no outbound traffic to the internal network was generated. In all breaches, the automated malware installed cryptominers in the bastion host, which explains why no interaction with the local file system nor private hosts was detected by \name. Since the GCP forbids crypto mining, we shut down the bastion host and restored it soon after each security breach. This can also explain the lack of interaction registered with the malware. It is possible that, given more time, the malware would have attempted to spread toward other hosts in the network to increase the mining capacity.

\subsubsection{Experimental Setup - Setting (II)}
The experimental setup for setting (II) consists of a single VM where we deploy \name equipped with two tools: Endlessh and Honeyports. Endlessh is configured to run on the standard port 22 in ``start'' mode, while we configure Honeyports to simulate several open services on well-known ports: FTP on 21, DNS on 53, HTTP server on 80, IMAP on 143 and MySQL on 3306.
The VM is deployed on the GCP and directly accessible on the Internet.

\subsubsection{Results - Setting (II)}
\begin{figure}
    \centering
    \includegraphics[width=0.8\columnwidth]{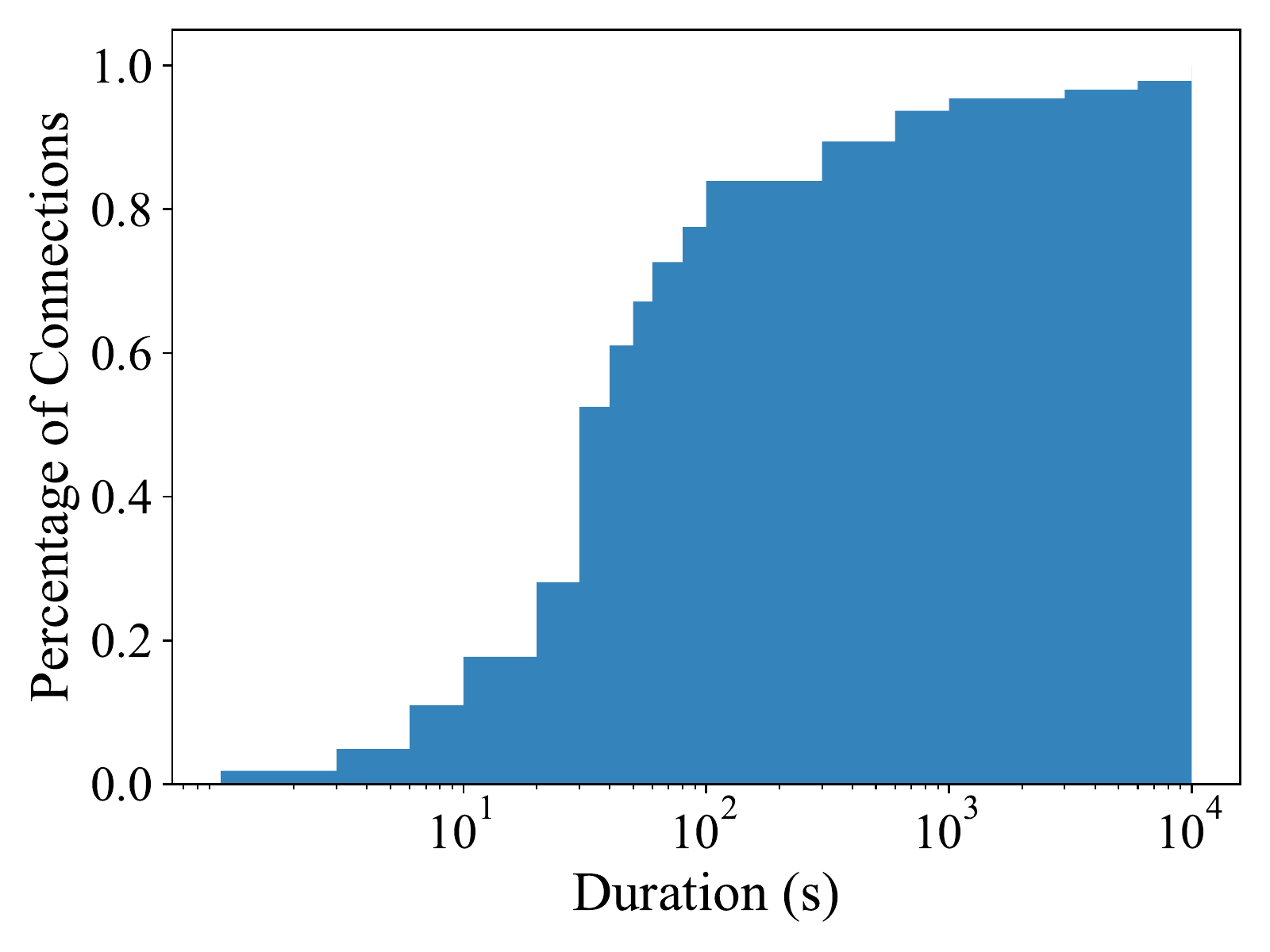}
    \caption{Cumulative Distribution Function (CDF) of Endlessh connections duration for the Setting (II) of the Automated Adversary evaluation.}
    \label{fig:ssh_duration}
\end{figure}

We have deployed the VM for a total of 430 hours, distributed over a period of one month. In total, five instances of the VM were deployed. Over this period, Honeyports detected a large number of incoming connections (port scans) evenly distributed among the different deployed services. More rarely, complete Nmap fingerprinting connections were detected as well --- also evenly distributed. Interestingly, the most significant portion of incoming connections did not target the Honeyports services, but concentrated on the ssh port: 1758 connections were detected by Endlessh over the 430 hours, with duration ranging from a few seconds to over 9 hours. This indicates a strong preference by the malware to target interactive services that can provide immediate access to the system, possibly because credential bruteforcing is an easily-automated process. Figure~\ref{fig:ssh_duration} shows the CDF of the inbound ssh connection duration registered during the evaluation. As we can see, most malware connections last for less than $100s$, with only a few long-lasting connections that skew the average. Most likely, the short-duration connections result from scanning/enumeration probes that give up after a short time, while longer-lasting ones correspond to real connections to attempt credentials bruteforcing.

\subsection{Lessons Learned: Automated Adversary}

\subsubsection{Lessons Learned - Setting (I)}
While it is unfortunate that only cryptominers interacted with our systems in the 30 days deployment window, the results of this experiment provide some interesting insights. Concerning research question (a), we can say that \name's file-level MTD modules are not helpful against cryptominers. This is unsurprising, as the only goal of this type of malware is to obtain as much computational power as possible to mine cryptocurrency. Indeed, all cryptominers in our experiments only downloaded scripts in the home directory, which is not considered suspicious activity.
The answer to research question (b) is also negative, as cryptominers do not appear to attempt any lateral movement once they obtain access to a machine. This result is somewhat surprising, as one would expect malware to try to infect as many devices as possible to further increase the available computational power. However, as we stated before, this result could be influenced by the short window of time the malware had to propagate to new machines before the bastion host was reset. Finally, our analysis of the filesystem-level activity performed by the cryptominers provides interesting hints on what new active defense tools can be developed against this type of malware. Table~\ref{tab:crypto_activities} details the activity carried out by the different cryptominers obtained through log analysis. Most of the activity carried out by the malware relates to downloading malicious scripts or unpacking malware in the \emph{/home} and \emph{/tmp} directories. Furthermore, at least one botnet exploits known services and cronjobs to ensure that the malware is running at all times. These interactions are challenging to deal with since introducing filesystem-level tools that detect the creation/modification of files in the \emph{/home} and \emph{/tmp} directories could hinder regular use of the system. However, almost all the malware detected also downloads additional scripts/bytecode beside the cryptominer to perform further actions. An interesting direction to deal with this kind of malware is to analyze the behavior these scripts enable and target these new actions with specific new deception MTB modules. Due to the terms of use of the GCP, we could not analyze the secondary behavior of these malware, as the bastion VM was terminated as soon as we detected crypto activity. The detailed analysis of the secondary behaviors of crypto miners and possible MTD countermeasures are left as future work.

\subsubsection{Lessons Learned - Setting (II)}
The results of this experiment show that \name's networking MTD modules are highly effective in hindering malware during the reconnaissance phase, forcing it to waste time and preventing it from being able to quickly probe all available services. Furthermore, the large amount of network noise generated makes it much easier to detect automated exploit attempts. The results of the experiments also suggest that malware often exploits easily automatable vulnerabilities, such as weak credentials. By using fake interactive services, such as Endlessh, to trap the malware in endless connection loops, real services can be protected from attacks. Therefore, our experimental results suggest a positive answer to the research question (c).

\subsection{Average Human Attacker}
\label{sec:eval_average}
This section evaluates the effectiveness of \name against average human attackers. For these experiments, we enrolled a team of students from the final year of our computer science degree and asked them to attack a VM protected by \name. The attackers aim to obtain local access to the VM and perform privilege escalation to steal sensitive documents. Similarly to our previous experiment, we design this evaluation to analyze the effectiveness of different \name MTD modules in deceiving and hindering knowledgeable adversaries. This experiment aims at providing a qualitative evaluation of \name and answering three research questions: (d) how effective are \name's networking MTD modules against human adversaries? (e) Can networking MTD tools effectively slow down and confuse human attackers, providing time for incident response teams to act? (f) Are filesystem MTD tools practical against privilege escalation and data stealing?

\subsubsection{Experimental Setup}
The experimental setup consists in one VM where we deploy real, vulnerable services as well as \name services. Attackers can interact with the VM remotely through the network. The detailed configuration of the VM is as follows:

\bpar{Real Services:}
\begin{itemize}
    \item Apache HTTP server on port 80. The server has a single webpage using CGI to handle user requests. The page shows a website under construction that internally calls a bash script to retrieve connection-related information to display. The bash version has been downgraded to be vulnerable to the Shellshock attack. The webserver runs under the default ``www-data'' user.
    \item SSH service on port 2000. This service is vulnerable to bruteforce attacks on two different user accounts using well-known weak passwords: ``test'' with password ``test''; ``user'' with password ``password123''. These users do not have root privileges.
\end{itemize}

\bpar{\name Services:}
\begin{itemize}
    \item Portspoof. Provides several fake services to slow down reconnaissance and make it harder to identify real services.
    \item Invisiport. Setup to trigger upon interaction with ports 53, 443, 445, corresponding to DNS service, HTTPS server, and Microsoft-ds (SMB - Server Message Block protocol). 
    \item Endlessh. Provides fake ssh tarpits on ports 22, 2001, and 2222.
    \item Honeyfile. Configured to log interactions with a set of directories that could be of interest to an attacker.
    \item Honeyfile. Configured to kill the PID and lock the user that accesses a set of directories hidden in the /home and /data directories.
    \item Cryptolocked. Configured to kill the PID and lock the user interacting with any trip file deployed in the machine's filesystem.
\end{itemize}
 
The pentesting activity was carried out in four separate sessions for a total of approximately 12 hours. The participants were not told about the presence of \name on the VM.

\subsubsection{Results}
During the initial phase of the attack, the pentesters carried out a port scan on the VM, detecting multiple open services. Initially, the pentesters focused mainly on port 445, the Server Message Block protocol, looking for misconfigurations that could potentially allow access to files or command execution. The pentesters also probed the service to enumerate potential system users. In reality, port 445 is a spoofed service provided by the Invisiport module, which automatically blocklisted the IP address of the pentesters, making all services unreachable except for a small set of fake services provided by the module itself. After further interaction, the pentesters noticed the inconsistencies caused by Invisiport and reset their session to restart the vulnerability assessment from scratch. In this second session, the pentesters focused mainly on the fake Samba and HTTPS services, primarily looking for ways to enumerate users and well-known web application vulnerabilities without success. During the third session, the pentesters switched focus to interactive services, particularly SSH. All four SSH services, one real and three fake SSH tarpits, saw considerable volumes of connections in this session. The fake SSH tarpit on port 22 absorbed the most significant number of connections, shifting the attention of the pentesters away from the real SSH service. Over the session, seven separate connections were directed to port 22, with an average duration of 30 seconds and a maximum duration of 115 seconds. Most of these were short-lived bruteforce attempts on the tarpit, which resulted in considerable wasted time. A small number of connections were also directed to the real SSH service and the other two SSH tarpits. However, no successful bruteforce attempt was registered.

In the fourth and final session of the test, we gave the pentesters local access to the machine by directing them to the real SSH service and suggesting bruteforcing with a list of well-known usernames/passwords. After obtaining local access, the pentesters began analyzing the filesystem looking for potentially valuable files and ways to escalate privileges. In their activity, the pentesters triggered some trip files, which resulted in Cryptolocked terminating the connection. The pentesters also used different scripts and tools to aid their search for potential vulnerabilities and valuable files, which resulted in numerous alerts from the deployed filesystem-based active defense tools. Finally, the pentesters identified a vulnerability in a Vim container, which allowed them to spawn a root shell and escalate privileges.

\subsection{Lessons Learned: Average Human Attacker}
From the logs gathered during the four pentesting sessions and a descriptive report produced by the pentesters, we can confidently say that the answer to research questions (d) and (e) is affirmative: \name networking MTD modules are highly effective at confusing and slowing down average human attackers. At the end of the third session, the testers reported that \textit{``the conflicting information gathered by initial assessments made it hard to identify attack vectors and plan exploitation strategies''} and that \textit{``active services that at times responded slowly and inconsistently complicated the pentesting process''}. Indeed, the Invisiport tool and SSH tarpits proved extremely valuable from the point of view of deception and annoyance, attracting many connections and absorbing a considerable amount of the pentester's time and attention. 

Based on the logs and alerts generated from the filesystem-level active defense tools, the answer to research question (f) is also affirmative: \name's filesystem MTD modules are a valuable asset for the early detection of privilege escalation attempts and data stealing. Trip files proved exceptionally effective against scripts aimed at gathering intelligence on the filesystem and system configuration-based vulnerabilities. The scripts used by the pentesters to detect privilege escalation vulnerabilities triggered multiple trip files and generated several log alerts that, in an entire enterprise network, would have immediately warned the CSIRT of the attack. 

Concluding, this evaluation shows that the coordination of multiple active defense tools by \name provides valuable defense in depth against average attackers. Moreover, it can significantly increase the time required for an attack thus amplifying the response time available to cybersecurity teams.

\subsection{Expert Human Attacker}
This section studies the effectiveness of \name against expert human attackers. We enrolled 31 final-year students with a solid background in systems security from our Master's degree program in Cybersecurity, and designed a multi-layer evaluation of our system. The goal of the attackers is to complete all layers of the experiment by compromising at least one VM in each layer and obtaining root privileges. The attackers had knowledge of the details of \name, but were not explicitly told of its presence on the VMs. We designed this evaluation to assess the effectiveness of \name as a deterrent and measure how much it can slow down attackers. To this end, layers one and three of the experiment are designed to be a one-to-one comparison between an attack on machines without \name and the same machines with \name. Layer two assesses the ability of \name to discourage attacks on systems by comparing a \name-protected VM to the same VM without \name-protection. This evaluation aims at answering two research questions: (g) is \name an effective deterrent? How much less likely to be compromised is a system running it? And (h) can \name provide additional time for incident response teams to react compared to systems without it? If so, how much?

\begin{table*}[t]
\footnotesize
\def\arraystretch{1.5}
\caption{Virtual machine configuration for the expert human attacker evaluation.}
\centering
\begin{tabular}[c]{| l | l | l | l | l |}

\hline
\textbf{Layer} & \textbf{VM} & \textbf{Service - Port} & \textbf{Vulnerabilities} & \textbf{\name Services - Port} \\
\hline
\multirow{5}{*}{Layer 1} 
 & VM1.1 & 
 \makecell[tl]{
		    \tabitem Bind DNS - 53 \\
		    \tabitem HTTP server - 80 \\
		    \tabitem Samba - 139, 445 \\
		    } & 
\makecell[tl]{
		    \tabitem SSH - 22 - remote user enumeration \\
		    \tabitem  FTP - 2121 - weak credentials \\
		    } & \tabitem  No \name  \\ 
\cline{2-5}
 & VM1.2  &
 \makecell[tl]{
		    \tabitem FTP - 21 \\
		    \tabitem Bind DNS - 53 \\
		    \tabitem Samba - 139, 445 \\
		    \tabitem Apache Tomcat - 80 \\
		    } & 
\makecell[tl]{
		    \tabitem SSH - 22 - remote user enumeration \\
		    \tabitem HTTP server - 8080 - command injection \\
		    } & \tabitem  No \name \\
\hline
\multirow{5}{*}{Layer 2} 
& VM2.1 & 
 \makecell[tl]{
            \tabitem SSH - 22 \\
		    \tabitem Bind DNS - 53 \\
		    \tabitem Rpcbind - 111 \\
            \tabitem MySQL - 3306 \\
		    } & 
\makecell[tl]{
            \tabitem HTTP  server - 80 - command injection \\
		    \tabitem Asterisk VoIP - 5038 -  PBX extension enum. \\
		    } & \tabitem  No \name  \\ 
\cline{2-5}
& VM2.2  & 
 \makecell[tl]{
		    \tabitem Bind DNS - 53 \\
		    } &
\makecell[tl]{
		    \tabitem SSH - 22 - remote user enumeration \\
		    \tabitem HTTP server - 3241 - command injection \\
		    } &
\makecell[tl]{
		    \tabitem Portspoof - multiple ports \\
		    \tabitem Endlessh - 2000, 2021 \\
            \tabitem Honeyfile \\
		    } \\
\hline
\multirow{5}{*}{Layer 3} 
& VM3.1 & 
 \makecell[tl]{
		    \tabitem Bind DNS - 53 \\
		    \tabitem HTTP server - 80 \\
		    \tabitem Samba - 139, 445 \\
		    } & 
\makecell[tl]{
		    \tabitem SSH - 2002 - remote user enumeration \\
		    \tabitem  FTP - 2321 - weak credentials \\
		    } &
\makecell[tl]{
		    \tabitem Portspoof - multiple ports \\
		    \tabitem Endlessh - 22 \\
            \tabitem Honeyfile \\
		    } \\
\cline{2-5}
& VM3.2  & 
\makecell[tl]{
		    \tabitem FTP - 21 \\
		    \tabitem Bind DNS - 53 \\
		    \tabitem Samba - 139, 445 \\
		    \tabitem Apache Tomcat - 8080 \\
		    } & 
\makecell[tl]{
		    \tabitem SSH - 2032 - remote user enumeration \\
		    \tabitem HTTP server - 3241 - command injection \\
		    } &
\makecell[tl]{
		    \tabitem Portspoof - multiple ports \\
		    \tabitem Endlessh - 22, 2000 \\
            \tabitem Honeyfile \\
		    } \\

\hline
\end{tabular}
\label{tab:vm_config}
\end{table*}

\subsubsection{Experimental Setup}
\begin{figure}
    \centering
    \includegraphics[width=0.9\columnwidth]{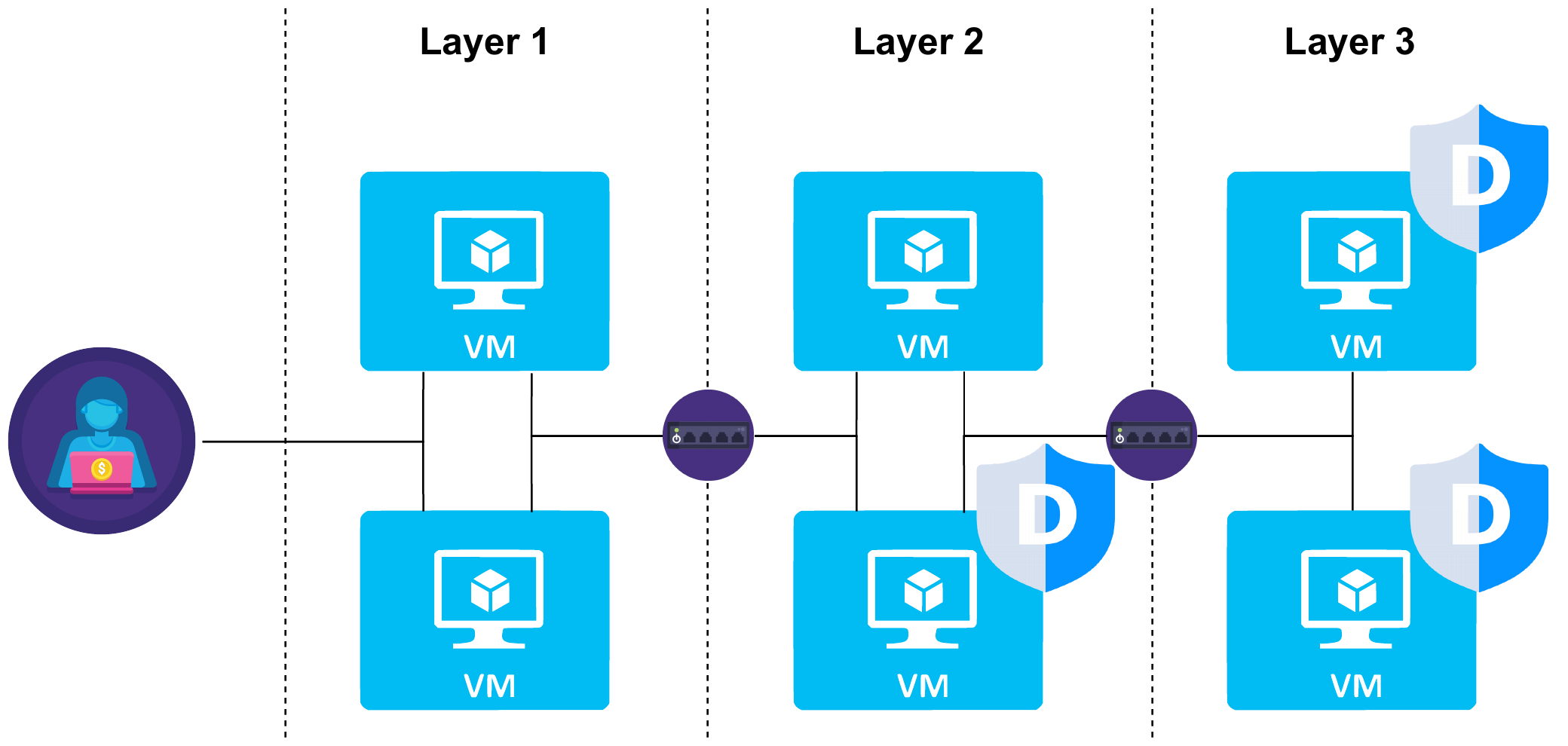}
    \caption{Experimental setup for the Expert Human Attacker evaluation.}
    \label{fig:expert_setup}
\end{figure}

The experimental setup used is illustrated in Figure~\ref{fig:expert_setup}. We use a three-layer design for this experiment where the attacker must compromise a VM in one layer before accessing the next one. Each layer comprises two VMs.
To unequivocally identify the VMs, we use the notation \emph{VMx.y}, where x identifies which layer the VM belongs to, and y identifies the VM number in that layer. For example, the second VM of the third layer would be VM3.2. The configuration of the VMs in each layer is presented in Table~\ref{tab:vm_config}. In the first layer, no machine uses \name, and both have reasonably straightforward vulnerabilities. In layer two, VM2.2 is protected by \name and has an easy-to-exploit command injection vulnerability on the webserver, similar to VM1.2. VM2.1, on the other hand, is not protected by \name but has a more complex exploitation path which requires successful PBX extension enumeration by exploiting the Asterisk VoIP service and then using one of these extensions to exploit a command injection vulnerability on a FreePBX web page~\cite{freepbx}. Finally, the third layer is configured with the same vulnerabilities as layer 1, but both machines are protected by \name.

\begin{table*}[t]
\footnotesize
\def\arraystretch{1.5}
\caption{Expert human attacker evaluation results. Shows the percentage of pentesters who successfully compromised each VM, the average time to compromise, and the overall traffic generated.}
\centering
\begin{tabular}[c]{| l | l | r | r | r |}
\hline
\textbf{Layer} & \makecell[c]{\textbf{VM}} & \textbf{Compromise \%} & \textbf{Time to Compromise} & \textbf{Traffic (GB)} \\
\hline
\multirow{2}{*}{Layer 1} 
 & VM1.1 & 5.34\% & 2h 1m 21s & 7.43 \\
\cline{2-5}
 & VM1.2 & 84.98\% & 1h 9m 36s & 13.79 \\
\hline
\multirow{2}{*}{Layer 2} 
 & VM2.1 & 61\% & 14h 37m 20s & 52.06 \\
\cline{2-5}
 & VM2.2 (\name) & 0\% & N.A. & 26.82 \\
\hline
\multirow{2}{*}{Layer 3} 
 & VM3.1 (\name) & 0\% & N.A. & 23.60 \\
\cline{2-5}
 & VM3.2 (\name) & 6\% & 48h 25m 42s & 31.29 \\
\hline
\end{tabular}
\label{tab:layered_results}
\end{table*}

\subsubsection{Results}
The results of our evaluation are presented in Table~\ref{tab:layered_results}. Overall, the pentesting activity lasted for over 57 hours over a period of one month and generated more than 150GB of traffic. More than 90\% of the attackers successfully compromised layer 1, with a significant preference for VM1.2 (84.98\% compromise rate). This VM was vulnerable to a reasonably direct command injection vulnerability, which could be found using the OWASP ZAP tool~\cite{zap}. VM1.1 saw considerably less interaction, almost half as much as VM1.2, and generally was ignored in favor of the other VM. From the penetration testing reports, no particular difficulties were reported on the first layer. This layer of the evaluation was designed to establish a baseline level of proficiency of the attackers and a baseline performance to compare \name's effectiveness. 

The second layer was successfully compromised by a considerably lower number of attackers (61\%), and \emph{all of them} achieved it by exploiting VM2.1, which did not run \name. As discussed, the exploitation path required to compromise this VM was much more complex than VM2.2, which had the same vulnerable web application as VM1.2. The fact that $~85\%$ of the attackers compromised VM1.2, but \emph{none of them} could exploit the same vulnerability in a VM with \name (VM2.2) unequivocally shows its effectiveness as a deterrent. We also note that many attackers reported suspicion that VM2.2 was a honeypot rather than a real system. These reports highlight an important advantage of embedding deception into production systems. 
This second layer demonstrates that the knowledge of the possible presence of deceptive tools in a system does not adversely impact the effectiveness of the deception. Indeed, it becomes an advantage for the defenders, as the attacker is led to exclude the easier VM in favor of the harder one, due to the fear that it might be a honeypot designed to deceive him. This behavior is consistent with recent findings in the literature on the effects of cyber deception on the behavior of attackers, which finds that ``The ambiguity effect suggests that ambiguity causes people to be unwilling to act'' and that being informed about the possible presence of deception can increase the confusion in the attacker~\cite{ferguson2021examining}, lending further credibility to our experimental results.

The final layer of the evaluation consists of just \name-protected systems. The VMs in the third layer were configured precisely like the VMs of layer 1, with only port numbers varying. From Table~\ref{tab:layered_results}, we see that only $6\%$ of the attackers successfully compromised this layer, compared to $90\%$ in layer 1. Moreover, the time-to-compromise for layer 3 was over 48 hours, compared to just $\sim1.5$ hours for layer 1: a 32-fold increase. The amount of generated traffic also dramatically increased between layer 1 and layer 3, more than doubling from 21.22GB in layer 1 to 54.89GB in layer 3. These results highlight the effectiveness of \name in defending production systems and significantly slowing down attacks, giving CSIRT ample time to take some actions. Furthermore, the considerable increase in registered network traffic  and the large number of alerts generated by \name (on average, 63 per attacker) eases the detection of intruders considerably.

\subsection{Lessons Learned: Expert Human Attacker}
From the results of this evaluation, we can draw several conclusions. We can claim that the answer to the research question (g) is affirmative. As shown by the layer 2 evaluation, \name is an effective deterrent, and machines protected by our system are unlikely to be targets of attacks. None of the expert attackers successfully compromised VM2.2, regardless of the easier exploitation path. Instead, after thoroughly interacting with it and generating over 26.82GB of traffic, they decided to avoid it, considering it a potential honeypot.

The answer to the research question (h) is also affirmative. The direct comparison between layers 1 and 3 shows that \name deception capabilities significantly increase system defenses and that, even when successful, expert human attackers require much more time and traffic to compromise the systems. The same VM configuration (VM1.2 - VM3.2) was much more challenging to exploit when \name was deployed because it decreased the success rate by $~93\%$ and increased the time-to-compromise by over 30 times.

In conclusion, the evaluation results show that \name is an effective defense tool against expert human attackers with considerable knowledge of system security.

\subsection{Professional Penetration Tester}
This section outlines the opinion that a professional penetration tester reported on \name. We provided the pentester with access to the architecture of \name and the MTD modules implemented and asked him how he would carry out an attack on a production system armed with \name.

The goal of this evaluation is to answer the following two research questions: (i) can \name be effective in slowing down professional attackers? (j) Can \name help CSIRT to expose stealthy attackers who try to compromise the systems without being detected?

We report the main conclusions of the professional pentester. He would begin the evaluation by obtaining information on the operating system run by the host, for instance through the \emph{ping} command. Based on the time-to-live field in the response packet, it is possible to identify the operating system running on the machine, as \name does not implement any MTD countermeasure against this type of interaction. Successively, the pentester would begin a system-wide scan to reveal open ports and an enumeration of running services with an automated tool. Since \name networking MTD modules are configured to block and prevent banner grabbing and enumeration, the results from the scan would appear suspicious to a professional. The pentester would then move to a more curated, stealth approach, enumerating the running services manually using a standard fixed request string. The majority of currently implemented \name networking modules provide a low-interaction environment designed to fool automated scanning tools. Through manual inspection, the professional pentester would slowly identify \name fake services and converge on the set of real services offered by the production system. During this process, the pentester's IP address would be blocklisted multiple times by the automated networking modules of \name, which would require a continuous IP change to pursue the attack. In the real world, this can be achieved by using multiple VMs from a cloud provider but would increase the time and cost required for exploitation. Eventually, the professional pentester would identify all \name fake services and obtain access to the production system by exploiting the real vulnerable services.

\subsection{Professional Penetration Tester: Lessons Learned}
\label{sec:eval_prof}
As is often the case with cybersecurity defenses, a professional adversary dedicating enough time, resources and effort would be able to identify \name MTD modules and eventually converge on the real production services. However, the goal of \name is not to stop an attacker completely but rather to slow him down and expose his activity to CSIRT for remediation. To identify the real production services, a professional attacker would have to manually interact with all available services and filter out \name's fake services one at a time. This activity requires a considerable amount of time, as it involves manual analysis of the responses from each deployed networking module. Moreover, interactions with several \name networking modules would result in blocklisting of the attacker's IP address. This would further increase the time required for the attack, forcing the adversary to switch between multiple machines to continue the activity. Finally, the interaction with \name networking modules would raise several alerts in the Controller which would relay them to the CSIRT. Therefore, we can conclude that the answers to research questions (i) and (j) are affirmative.

The professional pentester report provided us with important insights in how to improve \name and MTD techniques further. An important point highlighted by the pentester is that IP blocklisting makes it trivial to identify fake services. Indeed, interactions with spoofed services are often used as a trigger for IP blocklisting by \name. While this approach can be effective against automated tools and adversaries who rely on similar tools, this measure is not sufficient against a professional attacker, which can exploit this information to filter \name services. Furthermore, the professional pentester could filter out MTD modules individually because \name does not dynamically mutate the attack surface over time and the same port numbers stay assigned to the same fake services. The traditional MTD technique of periodically randomizing services would not work to address these issues. In fact, it would make it \emph{easier} for an attacker to identify real services: since \name would randomize only fake services, the remaining fixed, unchanging ones would be the real services. Allowing \name also to randomize real services would incur in all the drawbacks of traditional MTD randomization techniques, as well as providing \name with extended access to the real production system, which we strive to avoid. 

\vspace{0.4em}
\emph{Countermeasures.}
These limitations have been addressed and \name improved by integrating a new dynamic deception approach as described below. We can exploit the fake services exposed by \name to identify suspicious IP addresses and implement a \emph{transparent filter} on the open ports associated with real production services. Any IP address that interacts with a \name service is added to a suspicious list. Any connection coming from an IP in the suspicious list towards a real service is re-routed to a \name-controlled port where a similar fake service is exposed. From the adversary's point of view, the real service port is exposing a fake service. However, any legitimate user who did not interact with \name modules will be connected to the real production service without disruption. This dynamic deception approach can be further extended to introduce full, system-wide randomization of services without affecting any real service for legitimate users. After detection of suspicious activity from an IP address, \name can use transparent filtering to make real services unreachable for the attacker and at the same time re-randomize all running \name fake services. From the point of view of the adversary, this would effectively result in \emph{all} the services and ports on the machine changing at the same time. However, it would not affect legitimate users, and they would still be able to access real services without interruption. These functionalities do not require implementing new modules and are already achievable with the current \name architecture. The Management module of the \name Controller already implements a command to re-initialize all MTD modules deployed by an Agent, which can be used to provide re-randomization of the fake services. The implementation of the port filtering can be achieved by installing a simple rule in any programmable router or IDS in the network. The \name Controller instructs the relevant network device to change the destination port of any packet matching a specific triplet \emph{\textless source IP, dest. IP, dest. port\textgreater}, where source IP is an address in the suspicious list, and destination IP and port identify the socket of the real services to protect.

Finally, we can further strengthen the resiliency of the suspicious list by implementing \emph{enumeration fingerprinting} techniques in \name~\cite{kohno2005remote,nikiforakis2013cookieless}. The professional pentester report highlights how advanced adversaries employ manual techniques for enumeration to be more stealthy and to analyze the responses from the services better. Typically, an adversary does this through the use of fixed query requests using tools such as netcat. \name can exploit this by creating a fingerprint of all connection requests received by any of its networking modules. This fingerprint can be later used to identify further connection requests \emph{even if they come from the same adversary using other IP addresses}. We leave the implementation of the transparent filter module based on enumeration fingerprinting as future work.

\subsection{Overhead}
\label{sec:eval_overhead}
In the final section of our evaluation, we assess the impact of \name on system performance from the point of view of CPU and memory usage overhead. The main findings in Table~\ref{tab:overhead} show that the impact of \name on both CPU and RAM is negligible. When idle with only the core module loaded, the overhead in the system is essentially zero. Loading and deploying all \name tools results in similar negligible performance impact, with only $\sim11$MB of RAM used by the system. We could detect any measurable CPU overhead only when the system was actively under attack. When under port scan with all modules deployed, we registered a 0.1\% increase in CPU load and a $\sim41$MB overhead in system RAM usage. Similarly, with all modules deployed and a file-scanning script running in the system, we barely registered a 0.1\% increase in CPU usage and a 52MB increase in system RAM usage. Finally, when under both port scan and filesystem scanning, the CPU overhead reached a brief peek of 0.2\%, while the system RAM usage remained similar at $+52$MB. By all measures, we can say that \name adds negligible overhead to the systems in which it is deployed.

\begin{table}[t]
\footnotesize
\def\arraystretch{1.5}
\caption{Overhead of \name under different conditions.}
\centering
\begin{tabular}[c]{| l | r | r |}
\hline
\textbf{Configuration} & \textbf{CPU Overhead} & \textbf{RAM Overhead} \\
\hline
\name Core only & 0\% & 3.1MB \\
\hline
All modules idle & 0\% & 10.7MB \\
\hline
Under Port scan & 0.1\% & 40.8MB \\
\hline
Under filesystem scan & 0.1\% & 52.1MB \\
\hline
\makecell[cl]{
		    Under Port scan \\
		    Under filesystem scan \\
		    } & 0.2\% & 52.1MB \\
\hline
\end{tabular}
\label{tab:overhead}
\end{table}

\section{Discussion}
\label{sec:discussion}
This section discusses potential pitfalls of \name and future development directions toward a fully autonomous active defense agent.

\subsection{Increase in Attack Surface}
The main potential pitfall of \name is the increased attack surface it provides to adversaries. A general rule of cybersecurity is if you don't need something, then don't have it. This especially applies to open ports and exposed services. If a service is not necessary to achieve the function of a system, it should be removed, and its port should be closed to reduce the attack surface for potential adversaries. The \name approach of embedding fake services in production systems flips this concept on its head. While it is undeniable that adding fake services can potentially introduce new attack vectors in a system, we believe that our evaluation of \name shows the advantages outweigh the risks. The container-based architecture of \name and the multi-layer security features described in Section~\ref{sec:hardening} minimize the risk that compromise of \name services leads to privilege escalation and breach of the container. Moreover, by setting up seccomp profiles, capabilities, and AppArmor in a controlled manner, the container's access to the underlying system is restricted, reducing the ability of an attacker to exploit \name and compromise the production system.

\subsection{Towards an Autonomous Agent}
Currently, the \name Controller implements an interface through which security teams can view and update the configuration of \name modules, as well as view logging information and alerts. A promising future direction is to introduce a new component in the Controller that can autonomously deploy \name modules, as well as initialize their configuration based on the network environment and potential attackers. At the most basic level, such component could be modeled as an expert system with well-defined rules and conditions covering the most common use cases and system settings. A good starting point to design a system of this type would be the MITRE Engage framework~\cite{mitre_engage}, which defines a set of activities that can be performed in response to adversaries' actions. 

A more advanced, more flexible form of automation would be the introduction of an intelligent component in the Controller, able to make decisions based on the assessment of the state of the deployment environment and the actions taken by attackers. An interesting approach in this direction is deep reinforcement learning, which has been recently applied to many complex problems from robotics~\cite{nguyen2019review} to cybersecurity~\cite{nguyen2019deep}. Deep reinforcement learning is an interesting candidate for an intelligent controller since it excels at learning and solving complex high-dimensional problems. Controlling \name's module deployment and configuration while considering the current state of the environment and adversarial interactions is a problem with a very large state space and numerous potential actions that can be applied. Deep reinforcement learning~\cite{mnih2015human} uses an approximator, typically a deep neural network, to estimate the value function that is typically based on a table of state-action pairs in traditional reinforcement learning~\cite{qlearning}. The ability of the neural network approximator to learn a complex function mapping arbitrary states to action/rewards appears to be a perfect fit for the control problem of active defense tools. Some initial research on applications of reinforcement learning to active defense and cybersecurity tools configuration exists~\cite{eghtesad2020adversarial,klima2016markov,7417158}. However, these works are restricted to limited automation of specific tools, and no general solution to this problem exists to our knowledge. We leave the study of these research questions as future work.

\section{Related works}
\label{sec:related_works}
Researchers have proposed randomization and shuffling-based defense techniques for several decades, ranging from the first on n-version programming~\cite{chen1978n} to reconfigurable software and networks~\cite{compton2002reconfigurable}. However, this approach to cybersecurity was formalized only recently under the umbrella of Moving Target Defense~\cite{jajodia2011moving}. Since then, many approaches have been proposed in the literature at different levels of the system stack, with several review papers covering the topic \cite{TAN2023100544,9270287}. In this section, we analyze relevant related works and provide a comparison against \name.

\subsection{Hardware-level MTD Techniques}
In~\cite{kc2003countering}, the authors propose an MTD technique based on instruction-set randomization at the CPU level. They create process-specific randomized instruction sets based on a secret key  to defend against code-injection attacks. The idea is that an attacker who does not know the randomization key will inject code that is invalid for the specific randomized CPU, failing the attack. The authors of~\cite{portokalidis2011global} extend this approach to the whole software stack to prevent the execution of unauthorized binaries regardless of the attack vector. These approaches are orthogonal to our proposal, which does not aim to defend against this family of attacks.

\subsection{Network-level MTD Techniques}
Several works propose applications of different flavors of IP address randomization~\cite{al2013random,antonatos2005defending,sharma2018frvm,jafarian2012openflow}. Typically, this family of approaches periodically changes the IP address of production systems in a randomized manner to thwart adversarial reconnaissance.
The authors of~\cite{jafarian2012openflow} exploit Software-Defined Networking (SDN) to introduce an IP mutation technique that is transparent to end hosts. They exploit the global view provided by the SDN controller to keep track of two different IP for each host in the network: the real IP and an ephemeral ``virtual IP'' that changes at regular intervals.
In~\cite{sharma2018frvm}, the authors proposed an SDN-based IP multiplexing technique to randomize the addresses of end hosts in a similar fashion. Moreover, the authors of~\cite{yoon2020attack} propose a shuffling-based MTD technique that realies on SDN to randomize the network configuration of a host (mac/ip addresses, ports) based on the likelihood of it being in an adversary's attack path.
While promising, these proposals showcase the complications introduced by traditional MTD techniques: in order to utilize the defense, a network needs to use specific technologies (SDN in this case) and implement complex IP translation methods that complicate the maintenance of the network. \name avoids these complexities by introducing randomization only to fake services and resources, which greatly simplifies maintainability. In any case, both these approaches can be deployed alongside \name transparently, as \name does not affect real services in any way. In~\cite{al2013random}, the authors propose a similar IP mutation technique based only on routing updates. The proposal uses similar virtual/real IP pairs to randomly mutate the address of end hosts. At the network edges, real IPs are translated to virtual, which are used to route packets inside the network. The consistency of the routing is achieved by regularly updating the routing tables within the network with the current virtual IP for each host. Similarly to other IP-randomization proposals, this approach introduces high complexity in network management, with end hosts constantly changing IP addresses. Moreover, it also introduces additional overhead required for the convergence of routing paths after virtual IP updates, which increases proportionally to the size of the network and the average number of branches at each router.
Chowdhary et al~\cite{Chowdhary} propose another approach to counter DDoS attacks based on SDN, Snort IDS, and Nash Folk Theorem to analyze the behavior of suspected nodes.
The paper~\cite{kim2022divergence} introduces an MTD framework employing reinforcement learning to inspect network traffic and a network shuffling MTD technique to defend against detected threats. Similar to previous works, the proposal employs SDN to implement transparent IP randomization. These two approaches suffer the limitations mentioned above, namely the reliance on non-standard network technologies and increased network troubleshooting complexity.
In~\cite{10015577} Mani et al. analyze the resilience of address randomization-based MTD techniques. They show that machine learning-based can be employed to detect address randomization techniques and, in some cases, even to predict future assigned addresses. The analysis of the limitations of address-randomization techniques further adds to the limitations of these approaches.

\subsection{System-level MTD Techniques}
Another thread of MTD research focuses on system diversification and redundancy. This family of works employs multiple, diverse implementations of system components designed to achieve the same end result in different manners. In~\cite{zhu2013game}, the authors propose a theoretical model to fully randomize the network stack of systems. From all possible combinations of the different layers of the stack, a subset of allowed combinations is defined. The proposed system randomizes the layers periodically by choosing from this set of allowed combinations. The authors of~\cite{azab2011chameleonsoft} propose a system that changes components of a running program in a randomized manner. The authors use multiple functionally-equivalent implementations of different program components and randomly chose variants at runtime. In~\cite{gorbenko2009using}, the authors combine shuffling and diversity techniques and introduce a multi-version web service architecture designed to maximize system dependability. The paper~\cite{yuan2013architecture} introduces a redundancy-based MTD technique aimed at protecting webservers from command injection attacks. It uses multiple replicas of different software components of the webserver that are changed at runtime. 
All these works are orthogonal to our proposal and can be used in conjunction to \name to offer additional protection to production systems. However, it is worth noting that many of these techniques introduce considerable additional complexity in the management of systems, as well as sometimes prohibitive overhead~\cite{cho2020toward}. On the other hand, by limiting the application of MTD techniques only to deceptive services and system properties \name provides strong security at low complexity and overhead cost.

\subsection{Deception-based Techniques}
At a high level, deception-based techniques can be divided into two categories: (1) system-level deceptive techniques and (2) service-level deceptive techniques. System-level techniques provide complete, fake systems aimed at attracting the attacker's focus and slowing down his progress. Service-level techniques aim at altering the properties of existing services in deceptive ways to confuse adversaries. System-level techniques (1) are the most closely related to \name, since both are designed to limit changes to the production services, while at the same time slowing down attackers. Service-level techniques (2) are inherently orthogonal to our proposal, as they require changes to the real production services running on the systems, which is directly against one of the key design principles of \name. This section reviews both categories of deception based techniques. 

\subsubsection{System-Level Deception Techniques}
The authors of~\cite{ge2021proactive} propose a decoy-based shuffling approach to protect IoT devices. They exploit SDN to randomly perform network topology shuffling, hiding real IoT devices among fake decoy devices. 
In~\cite{9387290} the authors propose an adaptive cyber deception approach to predict the most likely attack path an adversary will follow, and successively deploy decoy nodes along the predicted path. 
Similarly to other deception-based approaches, these proposals suffer from the limitation that an adversary can typically detect decoy nodes with limited effort, and simply avoid them. \name avoids this drawback by embedding deceptive services directly into the production systems through the \name Agent, forcing attackers to either sift through numerous fake services to pinpoint the real ones, or drop the system entirely and move to a different target.
Recently Qin et al.~\cite{qin2023hybrid} provide a thorough evaluation on hybrid defense strategies relying on honeypots, honeynets, honeyfarms, honeytokens to provide cyberthreat intelligence and protect against attacks. Specifically, Honeynets~\cite{Zhang2010507, 7361266} consist of a virtual replica of a real-world production system that lacks production activities and authorized services. This deployment is primarily used for data collection, data which are further used in the detection and mitigation pipeline. Honeyfarms is a technique similar to honeynet used to simplify the deception network management that allows large-scale honeypot deployments. Similar to the above, these techniques suffer from limitations that they can be easily detected by the adversary due to the lack of production activities. Honeytokens~\cite{honeytokens1,honeytokens2} are another set of intrusion detection mechanisms that consist of the deployment of bait files. Honeytokens are similar to decoy files in working, which are already implemented in \name.

\subsubsection{Service-Level Deception Techniques}
In~\cite{araujo2014patches}, the authors propose a deception-based technique to defend systems against attacks exploiting known vulnerabilities. The core idea is that if a vulnerability is patched, an attacker exploiting it will immediately know because the attack fails.
The authors propose that security patches should include a deceptive component such that whenever an exploit against the patched vulnerability is attempted, the attack is transparently redirected to an unpatched decoy system, making the attacker believe the exploit succeeded. 
This approach is orthogonal to \name, which was designed with the specific goal of requiring no modification to the production systems, except for the deployment of the \name Agent. Furthermore, this approach requires manual activity to securely redact sensitive information before decoy generation. Nonetheless, the approach proposed by the authors can be adopted in conjunction with \name to increase the deceptive capabilities of the systems. 
Albanese et al~\cite{albanese2014manipulating} propose a graph-based approach to cyber deception aimed at altering the view an attacker has of a target system. The authors define a theoretical approach and algorithmic solution to maximize the distance between the attacker's view of a system and the real state of the system.
The main limitation of this proposal is that it requires modification to the real production systems and to the network traffic generated, which goes against the design principle of \name of not interfering with the behavior of real services in any way. 
In~\cite{gencc2019deception}, the authors analyze the effectiveness of decoy file-based defenses against cryptographic ransomware. 
Similarly, Ganfure et al.~\cite{10026355} study a similar decoy file-based defense called RTrap. Both these approaches focus on how to generate reliable decoy files that are hard to avoid by modern decoy-aware ransomware. These approaches are essentially an advanced version of traditional decoy files. \name already provides modules implementing decoy files, and RTrap's advanced functionalities can be easily incorporated into these modules.
In~\cite{tian2019moving} the authors propose a system-level approach to MTD to defend against Stuxnet-like malware, by including MTD in the control and sensing units of cyber-physical systems. The authors show the effectiveness of this approach through simulations of multiple attacks. This proposal is orthogonal to our approach, as \name is not specifically designed to run on resource-constrained cyber-physical systems. We plan to study the applicability of \name in these settings in future work.

\subsection{Autonomous Agents-based Techniques}
The family of works that are conceptually closest to \name is on autonomous cyber defense agents~\cite{kott2020doers,DeGaspari2019,kott2019autonomous,Theron2020}. Works in this family propose the use of agents based on artificial intelligence to reduce reliance on human cybersecurity experts and define a high-level architecture for the agents. These agents should be capable of autonomously planning and executing complex cyber defense activities to slow down and defeat automated malware. This family of works proposes only a high-level approximation of the architecture of such agents, and no implementation or detailed instantiation of the concept is given. 
In~\cite{sajid2021soda} the authors propose an orchestration framework, SODA, for cyber deception. SODA analyzes real-world malware behavior in the form of WinAPIs and maps them to MITRE ATT\&CK techniques. 
A set of deception-based techniques are then designed 
to deceive the specific threat detected in the systems. SODA relies on traditional detection agents to identify malware running on the system and builds a profile of the malware based on WinAPIs calls.
A subset of the suggested deception techniques is then selected by the user and applied through API hooking to deceive the malware. 
SODA has several differences and limitations compared to \name. The first, important difference is that SODA follows a traditional \emph{reactive} approach to cyber defense --- it requires an agent to detect the malware running ---. Reactive measures typically have poor effectiveness against quickly mutating threats, as adversaries can evolve their strategies to avoid static defenses. \name employs proactive techniques that are always active and don't rely on detection to thwart attackers. Secondly, SODA can provision systems only with limited deception capabilities at the API level, while \name can implement essentially any MTD/deception technique. Moreover, SODA implements deception capabilities as hooks in the underlying system, which requires tight integration with the production system. 
Third, SODA is limited in interacting only with malware that is \emph{already} running on the production system, since it requires analysis of the WinAPIs calls performed by the malware. This limitation heavily restricts its applicability compared to \name.
Finally, SODA is designed against static malware employing well-known attack patterns, and is not designed to defend against human adversaries. In contrast, \name is very effective in thwarting and slowing down even expert human attackers.
In~\cite{9999494} the authors propose an evolutionary approach to select the optimal MTD strategy based on a Wright-Fisher process. The proposed approach is designed to learn and describe the evolution trajectories of the attacker's and defender's strategies. This proposal is orthogonal to \name, as it focuses on the optimal orchestration of MTD tools. Nonetheless, this related work provides an interesting direction for future extensions of an intelligent DOLOS Controller.
\section{Conclusion}\label{sec:conclusions}
This paper presented \name, a novel architecture that unifies cyber deception and moving target defense approaches. \name is motivated by the insight that deceptive techniques are much more powerful when integrated into the production systems rather than deployed alongside them. \name combines typical MTD techniques such as randomization, diversity, and redundancy to deception, and brings all these functionalities directly into the production systems through the \name Agent.
We showed that \name can seamlessly and securely integrate into production systems through multiple layers of isolation, making it much more effective compared to typical deception techniques.
Through an extensive and thorough evaluation of \name against several different types of attackers, ranging from automated malware to professional penetration testers, we conclude that \name is highly effective in slowing down attackers and protecting the integrity of production systems.

Future work is needed to investigate the applicability of autonomous agents to the \name control logic for MTD modules deployment and orchestration. Furthermore, to provide a more complete context for autonomous decision-making, the Controller should correlate data coming from multiple Agents in real time. These improvements require a thorough comparison between synchronous and asynchronous communication paradigms with the controller, and an analysis of the advantages and limitations of each approach.

\section*{Acknowledgment}
This work was partially supported by project SERICS (PE00000014) under the NRRP MUR program funded by the EU-NextGenerationEU. This work was also supported by Gen4olive project funded by the European Union’s Horizon 2020 research and innovation programme under grant agreement No. 101000427.

\bibliographystyle{IEEEtran}
\bibliography{bibliography}

\end{document}